\pdfoutput=1
\documentclass[cernpreprint, atlasdraft=false, USenglish, texmf, orcidlogo]{atlasdoc}
\usepackage{atlaspackage}
\usepackage{atlasbiblatex}

\usepackage{atlasphysics}
\graphicspath{{logos/}{figures/}}

\usepackage{ANA-EXOT-2018-14-PAPER-defs}
\AtlasTitle{Search for quantum black hole production in lepton+jet
final states using proton-proton collisions at $\sqrt{s}$ = 13 TeV
with the ATLAS detector}

\AtlasVersion{2.6}

\AtlasAbstract{
A search for quantum black holes in electron+jet and muon+jet invariant mass spectra is performed with 140 fb$^{-1}$ of data collected by the ATLAS detector in proton-proton collisions at $\sqrt{s}$ = 13 TeV at the Large Hadron Collider. The observed invariant mass spectrum of lepton+jet pairs is consistent with Standard Model expectations. Upper limits are set at 95\% confidence level on the production cross-sections times branching fractions for quantum black holes decaying into a lepton and a quark in a search region with invariant mass above 2.0 TeV. The resulting quantum black hole lower mass threshold limit is 8.8 TeV in the Arkani-Hamed-Dimopoulos-Dvali model, and 6.3 TeV in the Randall-Sundrum model.
}

\author{ATLAS Collaboration}

\AtlasRefCode{EXOT-2018-14}

\AtlasDate{February 27, 2024}

\PreprintIdNumber{CERN-EP-2023-117}

\AtlasJournalRef{Phys. Rev. D. 109 (2024) 032010}
\AtlasDOI{10.1103/PhysRevD.109.032010}

\hypersetup{pdftitle={ATLAS document},pdfauthor={The ATLAS Collaboration}}

\begin{document}

\maketitle

%

%
\section{INTRODUCTION}
\label{sec:intro}

Quantum black holes (QBHs) are predicted in low-scale quantum gravity
models~\cite{ArkaniHamed:1998rs,Antoniadis:1998ig,Randall:1999ee} that offer
solutions to the mass hierarchy problem of the Standard Model (SM) by lowering
the scale of quantum gravity ($M_\mathrm{D}$) from the high Planck scale
($\sim$$10^{16}~\TeV$) to the \tev\ region (1--10~\tev).
In these new physics scenarios, gravity becomes strong, and quantum effects
are relevant. In models with large extra dimensions such as the
Arkani-Hamed-Dimopoulos-Dvali (ADD)
model~\cite{ArkaniHamed:1998rs,Antoniadis:1998ig}, the gravitational
field is allowed to propagate in $n$ extra dimensions ($n=6$ in our analysis),
while all SM fields are localized in the usual four-dimensional space-time.
There are also warped scenarios, such as the Randall-Sundrum model
(RS1)~\cite{Randall:1999ee}, in which a single warped extra
dimension ($n=1$) separates two three-dimensional branes by some distance.
Gravitons can propagate in this warped dimension, and the effective
Planck scale on the three-dimensional brane is determined by the
curvature of the extra dimension, also referred to as the warp
factor. These models postulate conservation of total angular momentum,
color, and electric charge in the production and in the decay of
QBHs~\cite{Meade08,Calmet08b,Gingrich:2009hj}. The behavior of QBHs
with masses near $M_\mathrm{D}$ decaying into two-particle final state
is distinct from that of the semiclassical black holes~\cite{Anchordoqui:2001cg}
that decay into a multiparticle final state via Hawking
radiation~\cite{EXOT-2015-09,EXOT-2015-04,CMS-EXO-15-007,CMS-EXO-17-023}
(thermal decay). Two-particle final states make up 51\% (74\%) of
all possible QBH decays in the ADD (RS1) model~\cite{Gingrich:2009hj}.

The QBH models can be tested at the Large Hadron Collider (LHC) up to 13~\tev.
In this paper, a search for QBHs decaying into a single electron ($e$) or
a single muon ($\mu$), and a quark producing a jet is undertaken.
The QBHs are postulated to be produced near the low-scale $M_\mathrm{D}$.
One expects in strong-gravity interactions that angular momentum, electric charge,
and color are conserved. It is less clear that global symmetries such as baryon or
lepton number of the SM need to be conserved in strong-gravity interactions.
While in high Planck-scale gravity in four dimensions the baryon number violation
is bound to be very small~\cite{Gingrich:2009hj}, the baryon number violation in
low-scale gravity in higher dimensions is less constrained and could cause a sizeable
impact on observables. Therefore, a search for QBH production that violates SM global
symmetries provides a possible way to examine low-scale gravity phenomena.
In the absence of a coherent and reliable Feynman diagram technique for the
quantum black hole description, the easiest and most accurate way to visualize
the QBH production mechanism would be a set of partonic 2-to-2 scattering processes
\begin{equation}
u u \to \bar{d} \ell^+,~~~u d \to \bar{u} \ell^+,~~~\bar{d} \bar{d} \to d \ell^+,
\label{eq:QBHproc}
\end{equation}
and the respective charge conjugates. Only these six electric charge initial states
($\pm$4/3, $\pm$2/3, $\pm$1/3) can result in a lepton-quark or lepton-antiquark pair
in the final state. Here, the $u$ and $d$ symbols denote all up and down quark flavors and
$\ell$---all charged leptons excluding $\tau$-lepton, which is not considered in the analysis.
In this way, all quark flavors are possible in both the initial and the final state in
Eq.~(\ref{eq:QBHproc}). The angular momentum of the QBH is entirely due to
the spin states of the incoming partons. The initial orbital angular momentum is assumed
to be negligible because of the small impact parameter in the parton-parton collision.
The model/generator makes no attempt to convert a classical impact parameter
to quantized orbital angular momentum.
Final states with lepton plus quark (antiquark) can only have either the spin 0 or
the spin 1. Thus, only QBHs with these spin states can contribute
to the cross section of the considered process.
Color is conserved in the interaction, but QBH has color, and thus the final state
has color so the beam remnant has the corresponding anticolor.

A previous search for QBHs in the lepton+jet channel was performed in proton-proton ($p$-$p$)
collisions at a center-of-mass energy of $\sqrt{s}=8~\TeV$ by ATLAS~\cite{EXOT-2013-03}.
The combined 95\% confidence level upper limit on the QBH production cross section with threshold mass
above 3.5~\tev\ was found to be 0.18 fb. This limit constrains the threshold mass of QBH,
which was found to be above 5.3~\tev\ in the ADD model. QBHs have also been sought in the
dijet, dilepton, and photon+jet channels by both ATLAS and CMS at center-of-mass energies of
7~\TeV~\cite{EXOT-2010-07,EXOT-2011-21,CMS-EXO-11-094},
8~\TeV~\cite{EXOT-2012-22,EXOT-2012-23,EXOT-2013-11,CMS-EXO-12-059,EXOT-2013-13}, and
13~\TeV~\cite{EXOT-2015-02,EXOT-2015-20,CMS-EXO-15-009,CMS-EXO-16-058,EXOT-2019-03,CMS-EXO-19-014,EXOT-2019-20}.
These LHC results with different final states use the same QBH model/generator, so results can be compared.
In general, the QBH searches in the lepton+jet final-state are less sensitive than in the dijet
searches (at the same QBH threshold mass). On the other hand, the limits obtained in the lepton+jet
events are stronger than those with photon+jet and dilepton final states.
Different final states supplement each other because they search for QBHs with different quantum numbers.
In addition, the lepton+jet model-independent limits can constrain some other models violating
the baryon-lepton global symmetry.


%
\section{ATLAS DETECTOR}
\label{sec:detector}
\newcommand{\AtlasCoordFootnote}{%
ATLAS uses a right-handed coordinate system with its origin at the nominal interaction point (IP)
in the centre of the detector and the \(z\)-axis along the beam pipe.
The \(x\)-axis points from the IP to the centre of the LHC ring,
and the \(y\)-axis points upwards.
Cylindrical coordinates \((r,\phi)\) are used in the transverse plane,
\(\phi\) being the azimuthal angle around the \(z\)-axis.
The pseudorapidity is defined in terms of the polar angle \(\theta\) as \(\eta = -\ln \tan(\theta/2)\).
Angular distance is measured in units of \(\Delta R \equiv \sqrt{(\Delta\eta)^{2} + (\Delta\phi)^{2}}\).}

The ATLAS experiment~\cite{PERF-2007-01} at the LHC is a multipurpose particle detector
with a forward-backward symmetric cylindrical geometry and a nearly \(4\pi\) coverage in
solid angle.\footnote{\AtlasCoordFootnote}
It consists of an tracking inner detector surrounded by a thin superconducting solenoid
providing a \qty{2}{\tesla} axial magnetic field, electromagnetic and hadron calorimeters,
and a muon spectrometer.
The tracking inner detector covers the pseudorapidity range \(|\eta| < 2.5\).
It consists of silicon pixel~\cite{PIX-2018-001}, silicon microstrip, and transition radiation tracking detectors.
Lead/liquid-argon (LAr) sampling calorimeters provide electromagnetic (EM) energy measurements
with high granularity.
A steel/scintillator-tile hadron calorimeter covers the central pseudorapidity range (\(|\eta| < 1.7\)).
The end cap and forward regions are instrumented with LAr calorimeters
for both the EM and hadronic energy measurements up to \(|\eta| = 4.9\).
The muon spectrometer surrounds the calorimeters and is based on
three large superconducting air-core toroidal magnets with eight coils each.
The field integral of the toroids ranges between \num{2.0} and \qty{6.0}{\tesla\metre}
across most of the detector.
The muon spectrometer includes a system of precision tracking chambers and fast detectors for triggering.
A two-level trigger system~\cite{TRIG-2016-01} is used to select events.
The first-level trigger is implemented in hardware and uses a subset of the detector information
to accept events at a rate below \qty{100}{\kHz}.
This is followed by a software-based trigger that
reduces the accepted event rate to \qty{1}{\kHz} on average
depending on the data-taking conditions.
An extensive software suite~\cite{ATL-SOFT-PUB-2021-001} is used in data simulation, in the reconstruction
and analysis of real and simulated data, in detector operations, and in the trigger and data acquisition
systems of the experiment.


%
\section{DATA SETS AND SIMULATED EVENT SAMPLES}
\label{sec:samples}

The results described in this paper use $p$-$p$ collision
data collected by ATLAS at $\sqrt{s} = 13$~\tev\
during 2015--2018 in stable beam conditions and with all detector
systems operating normally~\cite{DAPR-2018-01}. The event quality is checked to remove events
with noise bursts or coherent noise in the calorimeters.
Events in the electron channel are required to pass
at least one of two single-electron triggers~\cite{TRIG-2018-05}: the first
requires a transverse momentum (\pt) threshold
of 60~\GeV\ and the second has looser identification criteria and a \pt\ threshold
of 120 or 140~\GeV, depending on the data-taking period.
Events in the muon channel are recorded using a single-muon trigger~\cite{TRIG-2018-01}
with a transverse momentum (\pt) requirement of at least 50~\GeV.
The integrated luminosity of the dataset is determined to be $140.1\pm 1.2$~\ifb~\cite{DAPR-2021-01},
obtained using the LUCID-2 detector~\cite{LUCID2} for the primary luminosity measurements.

Background events with a high-\pT lepton and one or more jets arise from electroweak processes
including vector boson production with additional jets ($W/Z$+jets),
dibosons ($WW$, $WZ$ and $ZZ$), top-quark pair (\ttbar) and single-top-quark production,
and multijet processes including nonprompt leptons from
leptonic hadron decays and jets misidentified as leptons.

Monte Carlo (MC) simulation is used to model the expected contributions of various
SM processes as well as possible QBH signals. A full description of the
MC simulated event samples used is given below and summarized in Table~\ref{tab:MC}.
The expected contributions of the SM backgrounds reported in Table~\ref{tab:MC} are taken from
MC simulation, either directly or adjusted by the fit to data in dedicated control regions.
The multijet background is measured directly in data. In this case the events collected
by a set of unprescaled single-lepton triggers with different \pt-thresholds are used.

$W/Z$+jets and diboson samples~\cite{ATL-PHYS-PUB-2017-005,ATL-PHYS-PUB-2017-006}
are simulated with the \SHERPA generator~\cite{Bothmann:2019yzt}.
The $W/Z$+jets, and semileptonically decaying diboson samples, are simulated with \SHERPA 2.2.1, while
the fully leptonic diboson processes are simulated with \SHERPA 2.2.2.
In the \SHERPA samples the additional hard parton emissions~\cite{Gleisberg:2008fv}
are matched to parton showers based on Catani--Seymour dipole factorization~\cite{Schumann:2007mg}.
The \nnpdfnlo~\cite{Ball:2014uwa} set of parton distribution functions (PDFs) and a dedicated set of
tuned parameters developed by the \SHERPA authors are used~\cite{Bothmann:2019yzt}.
The matching of the matrix element to the parton
shower~\cite{Hoeche:2011fd,Hoeche:2012yf,Catani:2001cc,Hoeche:2009rj}
is employed for the various jet multiplicities, which are then merged
into an inclusive sample using an improved CKKW (Catani-Krauss-Kuhn-Webber)
procedure~\cite{Catani:2001cc} that is extended to next-to-leading-order (NLO)
accuracy using the MEPS@NLO prescription~\cite{Hoeche:2012yf}.
The virtual QCD correction for matrix elements at NLO accuracy is provided by
the \openloops\ library~\cite{Cascioli:2011va,Denner:2016kdg}.
The $W/Z$+jets (diboson) simulations are calculated for up to two (one) additional
partons at NLO and up to four (three) additional partons at Leading Order (LO). The $W/Z$+jets
processes are normalized to a next-to-next-to-leading-order (NNLO) cross section
prediction~\cite{Anastasiou:2003ds}. The diboson processes are normalized to the
NNLO cross section prediction~\cite{Grazzini:2015} as well.

The production of \ttbar~\cite{Frixione:2007nw} and single-top $tW$~\cite{Re:2010bp} and
\schan~\cite{Alioli:2009je} events is modeled using the
\POWHEGBOX~\cite{Nason:2004rx,Frixione:2007vw,Alioli:2010xd}~v2
generator at NLO with the \nnpdfnlo PDF set. The single-top
$t$-channel~\cite{Frederix:2012dh} is modeled with \POWHEGBOX
in the four-flavor scheme with the \texttt{NNPDF3.04fNLO} PDF set.
The events are interfaced with \PYTHIA 8.230~\cite{Sjostrand:2014zea} using the A14
tune~\cite{ATL-PHYS-PUB-2014-021} and the \nnpdftwo PDF set~\cite{Ball:2012cx}.
The \hdamp\ parameter\footnote{The \hdamp\ parameter controls the \pt\ of the first additional emission
beyond the leading-order Feynman diagram in the parton shower and therefore regulates the  high-\pt\ emission
against which the \ttbar\ system recoils.}
is set to 1.5 times the top-quark mass~\cite{ATL-PHYS-PUB-2016-020}.
The \ttbar\ inclusive production cross section is corrected to the theory prediction at
NNLO in QCD, including the resummation of next-to-next-to-leading logarithmic (NNLL)
soft-gluon terms calculated using \textsc{Top++2.0}~\cite{Czakon:2011xx}.
The $tW$ inclusive cross section is corrected to the theory prediction calculated at
NLO in QCD with NNLL soft-gluon corrections~\cite{Aliev:2010zk,Kant:2014oha}.
The \textsc{MadSpin}~\cite{Artoisenet:2012st} generator is used to preserve top-quark spin
correlations in the $t$-channel of the single top background. The \EVTGEN 1.6.0~\cite{Lange:2001uf}
package is applied for the modeling of c- and b-hadron decays.

\begin{table}[htb]
\centering
\caption{The event generators used for simulation of the signal and background processes.
The acronyms ME and PS stand for matrix element and parton shower.
The top-quark mass is set to 172.5 GeV.}
\label{tab:MC}
\footnotesize
\begin{tabular}{lll}\hline\hline
Process & ME generator and ME PDFs & PS, PDFs, nonperturbative effect \\\hline
$W/Z$+jets & \SHERPA 2.2.1, \nnpdfnlo & \SHERPA 2.2.1, \nnpdfnlo \\
$t\bar{t}$  & \POWHEGBOX, \nnpdfnlo & \PYTHIA 8.230, \nnpdftwo, \EVTGEN 1.6.0 \\
Single-top $s$-channel, $tW$& \POWHEGBOX, \nnpdfnlo & \PYTHIA 8.230, \nnpdftwo, \EVTGEN 1.6.0 \\
Single-top $t$-channel & \POWHEGBOX, \texttt{NNPDF3.04fNLO}, \textsc{MadSpin} & \PYTHIA 8.230, \nnpdftwo, \EVTGEN 1.6.0  \\
Diboson, semileptonic decay & \SHERPA 2.2.1, \nnpdfnlo &\SHERPA 2.2.1, \nnpdfnlo \\
Diboson, fully leptonic decay & \SHERPA 2.2.2, \nnpdfnlo &\SHERPA 2.2.2, \nnpdfnlo \\\hline
QBH signal, ADD, RS1 & \textsc{QBH} 3.0, CTEQ6L1 & \PYTHIA 8.205, CTEQ6L1, \EVTGEN 1.2.0 \\\hline\hline
\end{tabular}
\normalsize
\end{table}

The simulated QBH signal event samples are obtained from the \textsc{QBH} 3.0 generator~\cite{Gingrich:2009da},
which uses the CTEQ6L1 leading-order PDF set~\cite{ATL-PHYS-PUB-2014-021,Stump:2003yu}.
The parton showering and hadronization are performed in \PYTHIA 8.205, using
the CTEQ6L1 PDF set and the A14 tune. The QCD factorization scale for the PDFs
is set to the inverse gravitational radius~\cite{Gingrich:2009da}.
The QBH simulation assumes massless parton interaction and conserves total angular momentum.
The threshold mass, $M_\mathrm{th}$, is set equal to $M_\mathrm{D}$.
For the ADD QBH signal, the number of extra dimensions is $n=6$ (total number of dimensions $D=10$).
For the RS1 QBH signal, a single extra dimension is assumed leading to a total of five dimensions.
The ADD (RS1) samples for both leptonic channels are generated with $M_\mathrm{th}$ from
2~\tev\ to 9.5 (7.5)~\tev with steps of 0.5~\tev (the same as in Ref.~\cite{EXOT-2013-03}).
A quantum black hole is not a particle, so it does not have a single mass or width.
The generator produces a distribution of QBH masses (with no additional mass smearing).
The QBH mass is required to be in range of 1--3~$M_\mathrm{D}$ to ensure that the QBHs
are produced in the region in which expected quantum effects are important,
and that a region of possible thermal decay is excluded.
The decay products have exactly the energy and momentum of the produced black hole.
Unlike particles produced in quantum field theory, the black hole is produced in a
nonperturbative gravity model. The cross sections predicted by the \textsc{QBH} 3.0
event generator~\cite{Gingrich:2009da} are used in the determination of the model-dependent
limits for the signal processes. Processes with a quark pair in the initial state have
at least 2 orders of magnitude higher cross sections than those with antiquark pairs
in the initial state.

All simulated event samples include the effect of multiple $p$-$p$ interactions in the same
or neighboring bunch crossings. These effects are collectively referred to as pileup.
The simulation of pileup collisions is performed with \PYTHIA 8.186 using the ATLAS A3 set
of tuned parameters~\cite{ATL-PHYS-PUB-2016-017} and the \nnpdftwo PDF set
and weighted to reproduce the average number of pileup interactions per bunch crossing observed in data.
The generated background events are passed through a full detector simulation~\cite{SOFT-2010-01}
based on\ \GEANT~\cite{Agostinelli:2002hh}.
Simulated QBH event samples are produced with a fast parametrization of the calorimeter
response~\cite{ATL-PHYS-PUB-2010-013}, while \GEANT is used for the other detector systems.


%
\section{EVENT RECONSTRUCTION AND OBJECT IDENTIFICATION}
\label{sec:reco}

For an event to be considered, at least one $p$-$p$ interaction
vertex with at least two tracks must be reconstructed. The primary vertex is chosen to be the vertex with
the highest summed $\pt^2$ of tracks with transverse momentum $\pt>0.4$~\gev\ that are associated with
the vertex~\cite{ATL-PHYS-PUB-2015-026}.

Two identification levels are defined for leptons and jets, referred to as ``\texttt{Baseline}''
and ``\texttt{Signal},'' with \texttt{Signal} objects being a subset of \texttt{Baseline}.
The \texttt{Baseline} requirement provides a higher selection efficiency for leptons and jets
when calculating missing transverse momentum and resolving ambiguities between overlapping
physics objects (see below in this section). The leading lepton passing the \texttt{Signal}
selection is matched to the lepton that triggered the event.

Electron candidates are reconstructed using energy clusters in the EM calorimeter which
are matched to a track of Inner Detector (ID), and they are calibrated as described
in Ref.~\cite{EGAM-2018-01}. \texttt{Baseline} electron candidates are required to have
$|\eta|<2.47$ in order to pass through the fine-granularity region of the EM calorimeter
and be outside the range $1.37<|\eta|<1.52$ corresponding to the transition
region between the barrel and end cap EM calorimeters.
They should also satisfy \texttt{Loose} identification criteria
and have $\pt>10~\gev$. The trajectory of \texttt{Baseline} electrons must be consistent with
the primary vertex to suppress electrons originating from pileup.
Therefore, the tracks associated with \texttt{Baseline} electrons must have a longitudinal impact parameter
relative to the primary vertex ($z_0$) such that $|z_0\cdot\mathrm{sin}\theta|<0.5$~mm.
\texttt{Signal} electrons are defined as \texttt{Baseline} candidates that have $\pt>30~\gev$ and satisfy the
\texttt{Tight} identification and \textsc{HighPtCaloOnly} isolation requirements~\cite{EGAM-2018-01}.
The track associated with each \texttt{Signal} electron must have a transverse impact
parameter significance $|d_0 / \sigma(d_0)|\leq5$.

\texttt{Baseline} muon candidates are reconstructed in the region $|\eta|<2.5$
by matching ID tracks to tracks reconstructed in the Muon Spectrometer, and they are
calibrated \textit{in situ} using $Z\rightarrow \mu\mu$ decays~\cite{PERF-2015-10}.
\texttt{Baseline} muon candidates are required to have $\pt>10~\gev$.
They have to satisfy a set of requirements on the quality of the tracks defined as
\texttt{Medium}~\cite{PERF-2015-10} and to pass a requirement on the longitudinal
impact parameter $|z_0\cdot\mathrm{sin}\theta|<0.5$~mm.
\texttt{Signal} muons are defined as \texttt{Baseline} candidates that have $\pt>30~\gev$,
pass a requirement on the significance of transverse impact parameter $|d_0 / \sigma(d_0)|\leq3$,
and satisfy \texttt{HighPt} muon identification requirements~\cite{PERF-2015-10}
and a track-based isolation criterion. For the isolation requirement,
the summed \pt of tracks (with $\pt>0.4~\gev$) originating from the primary vertex
within a cone of radius $\Delta R=0.2$ around the muon, but excluding
the muon candidate track itself, has to be less than 1.25~\gev. A bad-muon veto for the
\texttt{HighPt} muons is applied. An event is rejected when \texttt{Signal} muon
has a large relative error of charge over momentum ($q/p$) associated with the track.
The veto requirement is changing from 2.5$\sigma$ to 2.0$\sigma$ depending on the muon
$\eta$ for the muon $\pt\le1~\tev$, and is linearly tightened to 1$\sigma$
for muons with $\pt\ge5~\tev$. The $\sigma$ is the average expected error
on the ($q/p$) as a function of the muon $\pT$ and $\eta$.

The \antikt algorithm~\cite{Cacciari:2008gp} with distance parameter $R=0.4$
implemented in the \texttt{FastJet} library~\cite{Fastjet}
is used to reconstruct jets up to $\abseta=4.9$
from massless clusters of energy depositions in the calorimeter~\cite{PERF-2014-07} (EMTopo jets).
Jets are then calibrated as described in Refs.~\cite{PERF-2014-02, PERF-2016-04}.
\texttt{Baseline} jets are required to have $\pt>20~\GeV$ and $\abseta<2.8$.
Events are vetoed if they contain jets induced by calorimeter noise or noncollision background,
according to criteria described in Ref.~\cite{ATLAS-CONF-2015-029}.
Additional jets that arise from pileup interactions are rejected by
applying a dedicated track-based selection (Jet Vertex Tagger~\cite{PERF-2014-03}),
based on classifying the tracks associated with the jet as pointing or not pointing to the primary vertex.
The jet candidates passing all the above requirements are called \texttt{Baseline} jets.
\texttt{Signal} jets are defined as \texttt{Baseline} candidates that have $\pt>30~\gev$.

Jets containing $b$-flavored hadrons, used only for the estimation of some backgrounds,
are identified in the region $\abseta<2.5$ by the MV2c10 algorithm~\cite{PERF-2016-05},
which makes use of the impact parameters of tracks associated with the candidate jet,
the positions of reconstructed secondary vertices and their consistency with the decay chains of such hadrons.
For the working point chosen for this analysis, such jets are identified with
an average efficiency of $77\%$ in simulated \ttbar\ events~\cite{FTAG-2018-01},
corresponding to rejection factors of 110, 4.9 and 15 for jets originating from
light quarks or gluons, charm quarks and $\tau$ leptons, respectively.

The efficiencies of the electron and muon trigger, reconstruction, identification and
isolation, the jet b-tagging and Jet Vertex Tagging, and the pileup rejection are taken
into account in every simulated event by applying the respective weights that correct for
deficiencies in the MC description of those efficiencies~\cite{EGAM-2018-01, PERF-2015-10,
PERF-2014-07, ATLAS-CONF-2015-029, PERF-2014-03, FTAG-2018-01}.

To avoid reconstruction of a single detected object as multiple leptons or jets, an overlap removal
procedure is applied to \texttt{Baseline} leptons and jets. First, jet candidates
are discarded if they are within $\Delta R<0.2$ of an electron. Second, electron candidates are
discarded within $\Delta R<0.4$ of the remaining jets. Finally, muon candidates are discarded
if they are within $\Delta R<0.4$ of a remaining jet with at least three tracks of $\pt>500~\mev$;
if this jet has less than three tracks, it is discarded and the muon is kept instead.

The missing transverse momentum (whose magnitude is denoted \met) is defined
as the negative vector sum of the transverse momenta of all identified objects
(electrons, photons, muons, jets and $\tau$-leptons)
and an additional soft term. The overlap removal between baseline objects is applied
before computing \met. The soft term is constructed from all tracks associated with
the primary vertex but not with any physics object. Fully calibrated electrons, muons,
photons, jets, hadronically decaying $\tau$-leptons and charged-particle tracks
are used to reconstruct \met~\cite{PERF-2016-07,ATLAS-CONF-2018-023}.


%
\section{EVENT SELECTION AND BACKGROUND ESTIMATION STRATEGY}
\label{sec:select}

The event selection is designed to be efficient for true electron+jet and muon+jet final states.
For candidate signal events, $\pt > 130$~\GeV\ is required for both the highest \pT\ (leading)
lepton and the highest \pT\ jet. The invariant mass of this lepton+jet pair, \Minv, is required to be
greater than 2.0~\TeV in the signal region. A veto on subleading leptons with $\pT>10~\GeV$ is applied.
Subleading jets in the event are required to have $\pT<130~\GeV$.
These selection requirements are summarized in Table~\ref{tab:select}.

\begin{table}[htb]
\centering
\caption{Definitions of the control, validation and signal regions.
The cells are kept empty in case the criterion is not applied.
Two same-flavor opposite-sign (SFOS) leptons satisfying the \texttt{Signal} selection
criteria are required in the $Z$+jets control and validation regions,
while \texttt{Signal} and \texttt{Baseline} stand for the corresponding sets of the lepton
and jet selection criteria.}
\label{tab:select}
\small
\begin{tabular}{lcccc}\hline\hline
Event selection               & WCR (WVR)           & ZCR (ZVR)           & TCR (TVR)             & SR (SVR) \\\hline
\Minv[TeV]                   & 1.0--1.5 (1.5--2.0) & 1.0--1.5 (1.5--2.0) & 1.0--1.5 (1.5--2.0)   & >2.0 (1.5--2.0) \\
Leading lepton, \pt[GeV]     & \texttt{Signal}, >130  & \texttt{Signal}, >130      & \texttt{Signal}, >130
& \texttt{Signal}, >130  \\
Subleading leptons, \pt[GeV] & \texttt{Baseline}, <10 & SFOS, >30      & \texttt{Baseline}, <10
& \texttt{Baseline}, <10 \\
Leading jet, \pt[GeV]        & \texttt{Signal}, >130  & \texttt{Signal}, >130      & \texttt{Signal}, >130
& \texttt{Signal}, >130  \\
Subleading jets, \pt[GeV]    & \texttt{Signal}, <130  & \texttt{Signal}, <130      & \texttt{Signal}, <130, N$\ge$3
& \texttt{Signal}, <130  \\
Number of b-tagged jets       &   0                 &                     &  $\ge$2               &  \\
\MET[GeV]                    & >60                 &                     &                       &  \\
$m_{\leplep}$ [GeV]           &                     & 70--110             &                       &  \\\hline\hline
\end{tabular}
\normalsize
\end{table}

The dominant background in both channels is the $W$+jets process in which the
$W$ boson decays leptonically. In the electron+jet channel, the second largest
background is events with nonprompt and misidentified (fake) leptons. It mostly originates
from multijet production processes when one of the jets is misidentified as a lepton.
In the muon channel, this background source is less than 0.5\% of the total background in the signal region (SR).
Its contribution is 4 times smaller than the single-top background and 10 times smaller
than the total uncertainty on the sum of all the other background contributions in the SR:
it is therefore considered to be negligible in the muon+jet channel.
There are also contributions from $Z$+jets events in which one lepton is not detected;
from diboson processes in which at least one boson decays leptonically;
as well as from \ttbar\ and single-top-quark production, in which the $W$ boson
from the top-quark decays leptonically.

The background yields for $W/Z$+jets and \ttbar processes in the SR are estimated
using dedicated control regions (CRs) and confirmed in validation regions (VRs).
The control (validation) regions enriched with $W/Z$+jets and \ttbar backgrounds are
designated as WCR (WVR), ZCR (ZVR) and TCR (TVR), respectively. They are orthogonal to each other.
There are different CRs and VRs in the electron+jet and muon+jet channels.
Definitions of all regions are given in Table~\ref{tab:select}.
The CRs/VRs are defined using \Minv\ requirements and additional selections
to increase the purity of the corresponding background (last three rows in Table~\ref{tab:select}).
The signal contamination estimated for the CRs is less than 0.3\%
for the ADD signal with $M_\mathrm{th}=5$~\TeV. This $M_\mathrm{th}$ value is considered
since lower masses were excluded by the previous analysis at 8~\TeV~\cite{EXOT-2013-03}.
An additional validation region, SVR, is used to verify the agreement of the background with data
in a phase space that is closer to the SR. The SVR uses the same selections as the SR
but with lower \Minv (see Table~\ref{tab:select}). %
The WVR and TVR are the subsets of SVR because extra requirements are applied to define
WVR and TVR selections in addition to the SVR selection. The ZVR is orthogonal to SVR
because two \texttt{Signal} leptons are required in ZVR.

The multijet background for the electron channel is estimated using the data-driven
\textit{Matrix Method} described in Ref.~\cite{EXOT-2018-30}.
Two parameters of the method (real and fake efficiencies, $r$ and $f$) are evaluated
using the MC simulated samples of the $W/Z$+jets background and the data samples.
Events in the samples are selected with looser object requirements with respect to the
\texttt{Baseline} selection to enrich the selected events with nonprompt electrons
and nonelectron objects identified as electrons.
The \textit{Matrix Method} uses \textit{tight} and \textit{loose} selection.
The \textit{tight} selection corresponds to the \texttt{Signal} requirement.
In contrast to \textit{tight}, the \textit{loose} selection uses the \texttt{Loose}
identification and does not apply the isolation requirement.

The $r$ value is the fraction of the electron candidates passing the \textit{tight}
requirements and matched to a generated electron, with respect to the electron candidates
passing \textit{loose} selection and matched also to a generated electron.
The $f$ value is the fraction of the electron candidates
passing the \textit{tight} requirements, but not matched to any generated electron (fake),
with respect to the candidate electrons passing the \textit{loose} selection and not matched to
any generated electron. The $r$ and $f$ efficiencies and their uncertainties are estimated
as a function of lepton $\pT$ and $\eta$, and they cover all regions.
The $r$ and $f$ efficiencies are a part of the fake/nonprompt lepton background's estimation
toolset~\cite{EGAM-2019-01} that is developed using data and MC simulations and is validated in data.
Estimation of the $r$ and $f$ uncertainties is described in Sec.~\ref{sec:syst}.
The number of events with fake electrons ($N_{\mathrm{multijet}}$) selected with the
\textit{tight} requirement is estimated as
\begin{equation}
N_{\mathrm{multijet}}=\frac{f}{r-f}\Bigl(r(N_l+N_t)-N_t\Bigr) ,
\label{eq:multijet}
\end{equation}
where $N_t$ is the total number of electron candidates passing the \textit{tight}
selection in the data sample. $N_l$ is the number of electron candidates that pass
the \textit{loose} selection and fail the \textit{tight} requirements in the data.

All background processes (and signal sources when they are included in the fit)
except the multijet are estimated using MC simulated events.
The control regions are used to constrain the freely floating $W$+jets, $Z$+jets and \ttbar\
background normalization factors, which are obtained independently for the electron and
muon channels. The normalizations for the multijet, diboson and single-top backgrounds
are allowed to vary, but only within their uncertainty ranges.


%
\section{STATISTICAL ANALYSIS}
\label{sec:stat}

A QBH signal is sought in the \Minv\ distributions
in the electron+jet and muon+jet channels as well as in their combination.
The statistical interpretation of the results is performed using
the profile likelihood method implemented in the HistFitter
framework~\cite{Baak:2014wma}. The likelihood function
is a product of the probability density functions of the
binned \Minv\ distributions, with one for each region contributing to the fit.
The number of events in each of the bins in the given
regions is described using a Poisson distribution, the mean of which is the
sum of the expected contributions from all background and signal sources.
Systematic uncertainties described in Sec.~\ref{sec:syst} are added into the fit as
nuisance parameters. They are assumed to follow Gaussian distributions
whose widths are determined from the size of the corresponding uncertainty.
Normalization factors are free-floating parameters in the fit.
All the fit parameters are determined by maximizing the product of the Poisson
probability functions and the constraints on the nuisance parameters.

The combination of the electron and muon channels was made by merging the
electron and muon samples in the data and in the MC simulation.
The combined channel (lepton+jet) is fitted independently from the electron
and muon channels. Two types of fits are performed as detailed below.

A \textit{model-independent} fit compares the data event yield in the SR with
the SM background estimate and its uncertainties, to test for possible contribution of any
non-SM signal in the SR. As a first step, a \textit{background-only} fit is performed,
where the normalization and shape fit of the backgrounds is adjusted to match the data
in the three control regions simultaneously. The resulting distributions are extrapolated into
the signal region to correct the expected shapes and yields of the corresponding backgrounds.
The extrapolation of the adjusted distributions and nuisance parameters is also checked
in the VRs by means of comparison to data and total yield of the SM background. In a second step,
any non-SM signal is sought in the SR. The possible contribution of a signal is scaled by
a freely floating normalization factor of the dummy signal added in the SR. The significance of
a possible excess of observed events over the SM prediction is quantified by the one-sided
probability, $p_0$, of the background alone to fluctuate to the observed number of events or
higher, by using the asymptotic formula described in Ref.~\cite{Cowan:2010js}. The presence of
a non-SM signal would manifest itself as a small $p_0$ value. In the absence of an excess
over the SM expectation, upper limits on the cross section of any non-SM signal are estimated.

In a \textit{model-dependent} fit, an ADD or RS1 signal is included in the SR,
and its yield is scaled by a freely floating signal normalization factor.
In the absence of any significant excess above the SM background prediction, limits are
evaluated with the modified frequentist CL$_\mathrm{S}$ method~\cite{Read:2002hq}.
The background normalization factors and nuisance parameters are determined simultaneously
in the CRs and in the SR. The bin width over \Minv in the SR is optimized to obtain
good fit performance and stability for all QBH threshold masses used in the analysis.
The 2 TeV width was found to be the best bin size.

Acceptance and efficiency are estimated with the use of the simulated QBH
signal event samples. Acceptance is calculated at the generation level
as the fraction of events passing the signal requirements.
Efficiency is the fraction of events passing the signal requirements at the
reconstruction level with respect to the generation-level signal requirements.
The product of acceptance and efficiency (\textit{Acc}~$\times$~\textit{Eff})
of the signal selection does not depend on the QBH threshold mass
within their uncertainties. The \textit{Acc}~$\times$~\textit{Eff} is also consistent
for both models (ADD and RS1). The average
values of \textit{Acc}~$\times$~\textit{Eff} are equal to $(66.5\pm 0.4)\%$ and
$(67.1\pm 0.4)\%$ in the electron and muon channels, respectively.

The suppression of the additional jet activity in the event due to the vetoing of
subleading jets with $\pT>130~\GeV$ leads to a better separation between the signal
and SM background processes. However, the constraint distorts the acceptance and efficiency
of the signal extraction from the background, since the QBH signal is calculated at LO+PS
accuracy in QCD, while the largest SM backgrounds, $V$+jets, are generated with NLO+PS precision.
Thus, the comparison of signal with background may be distorted in the fit, leading to an
over-optimistic estimate of upper limits on $\sigma\times Br$. The effect of the higher order
QCD radiation in the QBH production yield is evaluated as a correction factor, $R_c$.
The $R_c$ is obtained using the $W/Z$+jets samples, since the events have a color structure
in the final state similar to that of the signal and are generated at higher order accuracy.
Hence, the $R_c$ quantifies the overestimate in signal acceptance and efficiency due to
use of the veto on the high-\pT subleading jets at LO MC generated events.

The $R_c$ is defined as the ratio of the number of events passing the signal selection
without and with the requirement on the subleading jet activity (see Table~\ref{tab:select}).
The ratios are calculated separately for $W$+jets and $Z$+jets events, and the average
of the two is used as the $R_c$ correction factor. The maximal difference between corrections
obtained with the $W$+jets and $Z$+jets samples is used as the systematic uncertainty on the $R_c$
factor. Statistical uncertainties in the $W/Z$+jets samples are also included in the $R_c$
total uncertainty. The uncertainty on $R_c$ was added to the total systematic uncertainty in the fit.
The QCD correction for the electron+jet and muon+jet final-states combination is calculated
as the weighted average. The $R_c$ correction factors in the electron and muon channels and their
combination are given in Eq.~(\ref{eq:R_cor}), respectively:
\begin{equation}
\left<R_c\right>^\mathrm{ele}=0.36\pm0.02,~~~\left<R_c\right>^\mathrm{muo}=0.39\pm0.04,~~~\left<R_c\right>^\mathrm{comb}=0.37\pm0.02.
\label{eq:R_cor}
\end{equation}
The $R_c$ factors are consistent between the electron, muon and combined channels and are used
to correct the signal MC \textit{Acc}~$\times$~\textit{Eff} in the limit setting procedure.


%
\section{SYSTEMATIC UNCERTAINTIES}
\label{sec:syst}

Systematic uncertainties are evaluated for all signal and background predictions
and include experimental uncertainties on detector measurements as well as
modeling uncertainties and the effect of limited statistics of MC simulation.
The systematic uncertainties of all backgrounds are extrapolated from the control regions
into the validation and signal regions in the background-only fit.
The background uncertainties are practically the same in all types of fit.
The expected QBH signal and its uncertainties are estimated for the ADD and RS1 models
in the model-dependent fit. The relative systematic uncertainties for the SM background and
a representative signal (ADD, $M_\mathrm{th}=6.0$~\tev) in the SR are represented in
Table~\ref{tab:syst}. The resulting uncertainty in the total background differs from
the sum in quadrature of the single sources because of correlations.

\begin{table}[htb]
\centering
\caption{The relative systematic uncertainties (in \%) on the
SM background in the SR are estimated in the background-only fit; and systematic uncertainties
on the ADD signal are estimated for the QBH with $M_\mathrm{th}$ = 6.0 TeV
in the model-dependent fit. Lepton modeling combines all the types of experimental
uncertainties for the electrons or muons. All the uncertainties shown are
obtained independently for the electron and muon channels.
The relative statistical errors on the data (in \%) are also shown. }
\label{tab:syst}
\begin{tabular}{lcccc}\hline\hline
Source & \multicolumn{2}{c}{Electron+jet} & \multicolumn{2}{c}{Muon+jet} \\
& Background & Signal & Background & Signal \\\hline
JER                        & 2.4 & 1.9 & 2.4 & 1.6 \\
JES                        & 0.7 & 0.4 & 0.6 & 0.5 \\
Lepton modeling            & 2.8 & 0.6 & 3.6 & 1.7 \\
Pileup                     & 0.7 & 0.6 & 0.8 & 1.0 \\
Luminosity                 & 0.5 & 0.7 & 0.5 & 0.7 \\
$W$+jets normalization     & 1.1 &     & 1.1 &     \\
$W$+jets modeling          & 0.5 &     & 0.6 &     \\
$Z$+jets normalization     & 0.3 &     & 0.3 &     \\
$Z$+jets modeling          & 0.3 &     & 0.3 &     \\
\ttbar\ normalization      & 0.2 &     & 0.4 &     \\
MC statistics              & 1.6 & 0.6 & 1.5 & 0.7 \\
Multijet estimation        & 1.4 &     &     &     \\\hline
Total uncertainty          & 4.6 & 2.4 & 5.1 & 2.7 \\\hline
Statistical errors of data & 2.1 &     & 2.7 &     \\\hline\hline
\end{tabular}
\end{table}

Experimental uncertainties reflect the accuracy of the experimental measurements of
jets and leptons. The jet energy scale (JES) and resolution (JER) uncertainties
are derived as a function of the \pt\ and $\eta$ of the jet. They are determined
using a combination of data and simulation, through measurements of the jet \pt balance
in dijet, $Z$+jets and $\gamma$+jets events~\cite{PERF-2016-04}. The uncertainties in scale
and resolution of the electron energy~\cite{EGAM-2018-01} and muon momentum~\cite{PERF-2015-10}
are propagated to the measured event yield. Systematic uncertainties in the measurements
of the electron~\cite{EGAM-2018-01,TRIG-2018-05} and muon~\cite{PERF-2015-10} identification,
reconstruction, isolation, and triggering efficiencies as well as in the pileup jet identification
using the jet vertex tagger algorithm~\cite{PERF-2014-03} are also propagated to the measured
\Minv distributions.

The uncertainty in the \Minv spectrum due to pileup is estimated by varying the average number
of pileup events in the simulation to account for the differences between the values
of the measured and predicted total inelastic cross section used in the pileup simulation~\cite{STDM-2015-05}.
The impact of the luminosity uncertainty on the SM background is estimated
by varying the integrated luminosity combined over 2015--2018 within its
uncertainty of 0.83$\%$~\cite{DAPR-2021-01}.

Modeling uncertainties on the $W/Z$+jets backgrounds are calculated as follows.
The PDF uncertainties propagated to the \Minv distribution are estimated
using the nominal PDF set and a set of 100 PDF replicas for \nnpdfnnlo~\cite{Ball:2014uwa}.
The impact of the $\alpha_s(m_Z)$ uncertainty on the background is estimated
by varying $\Delta\alpha_s(m_Z)=\pm 0.002$.
The impact of missing higher order calculations is evaluated using seven-point variations
of the factorization and renormalization scales in the cross section calculations.
The scales are independently varied upward and downward by a factor of 2, excluding
simultaneous variations in opposite directions. The envelope of the resulting variations
as a function of \Minv\ is taken as the size of the associated systematic uncertainty.
All aforementioned modeling uncertainties are combined in quadrature and represented
in Table~\ref{tab:syst} as ``$W/Z$+jets modeling.'' Total modeling (theoretical) uncertainties
are not estimated for the \ttbar, single-top and diboson samples because they are small backgrounds.
The uncertainties in the normalization of the $W/Z$+jets and \ttbar\ backgrounds from the
fitting procedure are shown in Table~\ref{tab:syst} as well as uncertainties from the
limited MC statistics of the background simulated samples.

The uncertainties in the multijet background are related to the estimate of the $f$ and $r$
parameters ($\Delta f$ and $\Delta r$) as well as to statistical errors in the total number
$N_t$ of \textit{tight} electron candidates ($\Delta N_t$) and in the total number $N_l$ of
\textit{loose} electron candidates ($\Delta N_l$). The $\Delta f$ and $\Delta r$ uncertainties
are estimated for different ($\eta-\pT$) regions by varying the requirements used in the event
selection~\cite{EXOT-2018-30}. All these uncertainties are combined in quadrature and reported
in Table~\ref{tab:syst}.

Systematic uncertainties described in this section are added into the fit as nuisance parameters
where they can be pulled and constrained. After the fit all systematic uncertainties are pulled
by less than 0.7 of a standard deviation. The errors of nuisance parameters are constrained
within $\pm0.5\sigma$ in comparison with their initial values.


%
\section{RESULTS}
\label{sec:result}

In the \textit{background-only} fit, the normalization factors of the $W$+jets, $Z$+jets
and \ttbar background processes are consistent with unity within uncertainties.
Differences of normalization factors from unity are $\le$5\% in all cases.
The \Minv distributions of events in the WCR, ZCR, TCR and corresponding validation regions
after the \textit{background-only} fit are shown in Fig.~\ref{fig1}. There is good agreement
between the data and the SM background in all CRs and VRs.

The comparison of the post-fit background yields with the data in the SVR and SR
is represented in Table~\ref{tab:yields}. The pre-fit background yields expected
in the MC are shown in the bottom part of Table~\ref{tab:yields}. There is agreement
between the data (``Observed data'') and the total SM background (``Fitted events'')
within 1$\sigma$ in all regions. The errors include both statistical and systematic uncertainties.

\begin{table}[htb]
\centering
\caption{The observed number of data events, the fitted
background events in the SVR and the SR for the \textit{background-only} fit
and the number of background events expected from
the MC background samples in the electron+jet and muon+jet channels.
The errors shown for the ``Expected events'' are statistical
and systematic uncertainties summed in quadrature.}
\label{tab:yields}
\begin{tabular}{lrrrr}\hline\hline
& SVR electron+jet & SVR muon+jet   & SR electron+jet & SR muon+jet    \\\hline
Observed data   & $9053$           & $5504$         & $2319$          & $1359$         \\\hline
Fitted events   & $8900 \pm 320$   & $5380 \pm 200$ & $2290 \pm 110$  & $1386 \pm 70$  \\\hline
$W$+jets   & $5590 \pm 270$   & $4190 \pm 200$ & $1290 \pm 70$   & $1087 \pm 54$  \\
Multijet   & $1670 \pm 200$   &                & $570 \pm 47$    &                \\
$Z$+jets   & $646  \pm 73$    & $439  \pm 27$  & $199  \pm 17$   & $131  \pm 13$  \\
\ttbar\    & $527  \pm 10$    & $351  \pm 7$   & $109  \pm 5$    & $69   \pm 5$   \\
Single top & $143  \pm 7$     & $112  \pm 5$   & $31   \pm 2$    & $28   \pm 2$   \\
Dibosons   & $335  \pm 22$    & $289  \pm 14$  & $94   \pm 9$    & $72   \pm 8$   \\\hline
Expected events & $9390 \pm 340$   & $5260 \pm 220$ & $2647 \pm 94$   & $1303 \pm 55$  \\\hline
$W$+jets   & $6090 \pm 270$   & $4080 \pm 210$ & $1654 \pm 65$   & $1016 \pm 48$  \\
Multijet   & $1690 \pm 210$   &                & $577 \pm 38$    &                \\
$Z$+jets   & $598  \pm 85$    & $408  \pm 23$  & $186  \pm 18$   & $122  \pm 12$  \\
\ttbar\    & $546  \pm 14$    & $366  \pm 7$   & $109  \pm 6$    & $71   \pm 5$   \\
Single top & $141  \pm 7$     & $104  \pm 4$   & $29   \pm 2$    & $28   \pm 2$   \\
Dibosons   & $327  \pm 23$    & $298  \pm 12$  & $92   \pm 10$   & $66   \pm 8$   \\\hline\hline
\end{tabular}
\end{table}

The difference between expected and fitted yields of $W$+jet background in the
electron channel in Table~\ref{tab:yields} is $\sim$10\% in SVR and $\sim$20\% in SR,
with the $W$+jet normalization factor close to unity ($1.01\pm0.02$), since
the pre-fit $W$+jet background in the electron channel has
a visible slope relative to data. The $W$+jet distribution is fitted to data
in CRs where both the slope and the normalization are adjusted. The slope elimination
is a result of simultaneous pulls of several nuisance parameters. The main contributors
are five parameters that tune the $W$+jet yields in the 5-bin WCR in the likelihood fit.
The total uncertainty includes statistical errors, uncertainties on normalization and
slope elimination, and nuisance parameters related to objects, detector, and modeling.

\begin{figure}[!htbp]
\centering
\subfloat[]{\includegraphics[width= 0.4\textwidth]{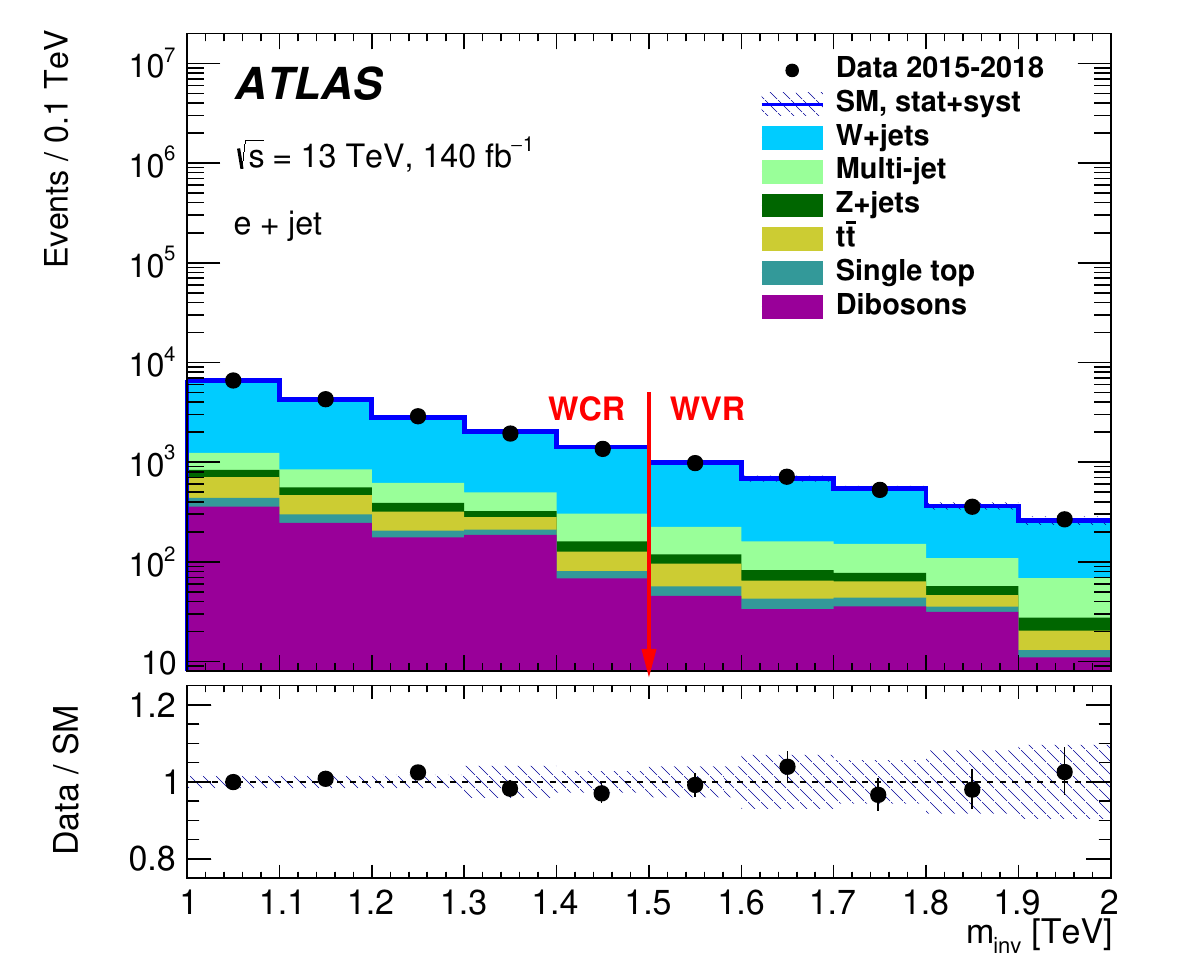}}
\subfloat[]{\includegraphics[width= 0.4\textwidth]{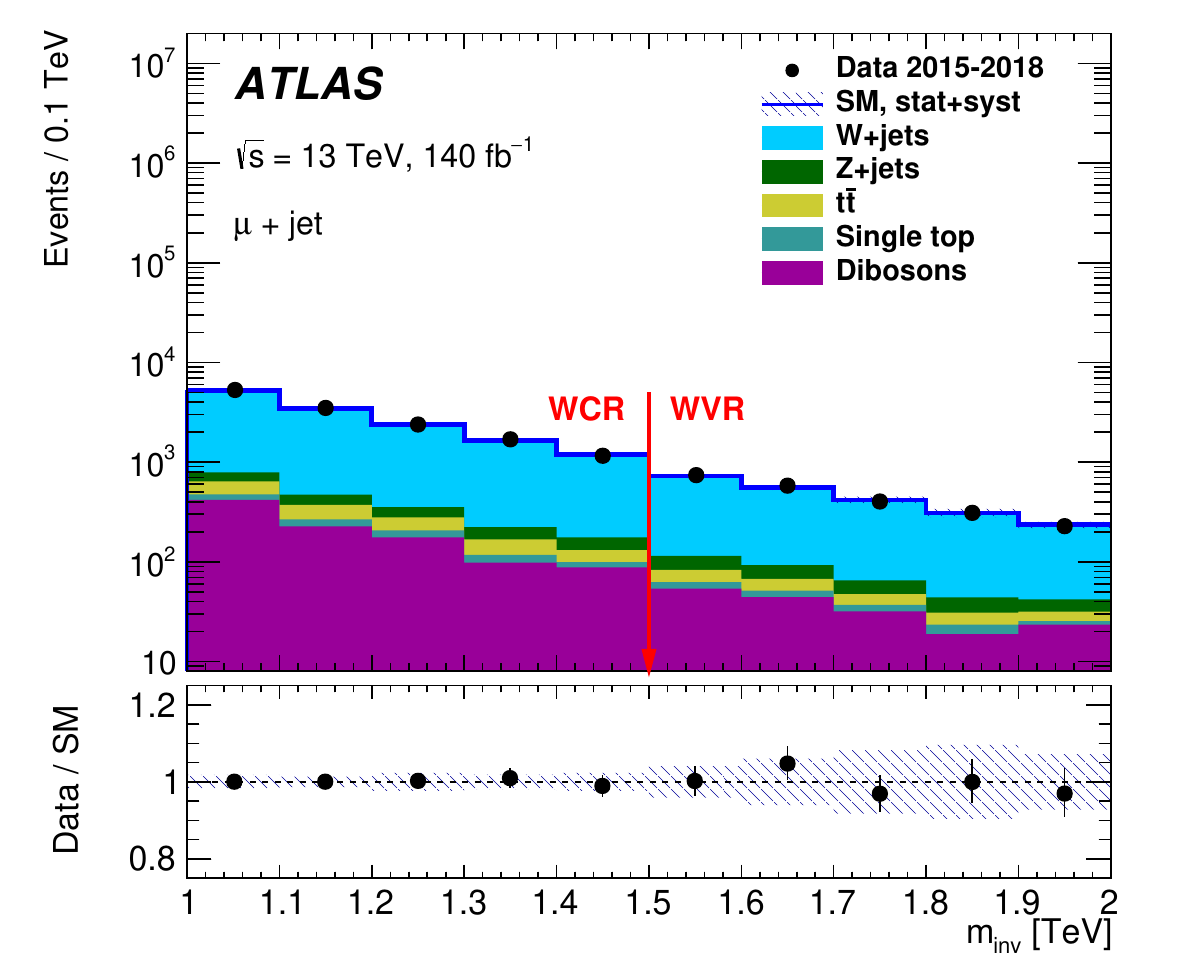}}\\
\subfloat[]{\includegraphics[width= 0.4\textwidth]{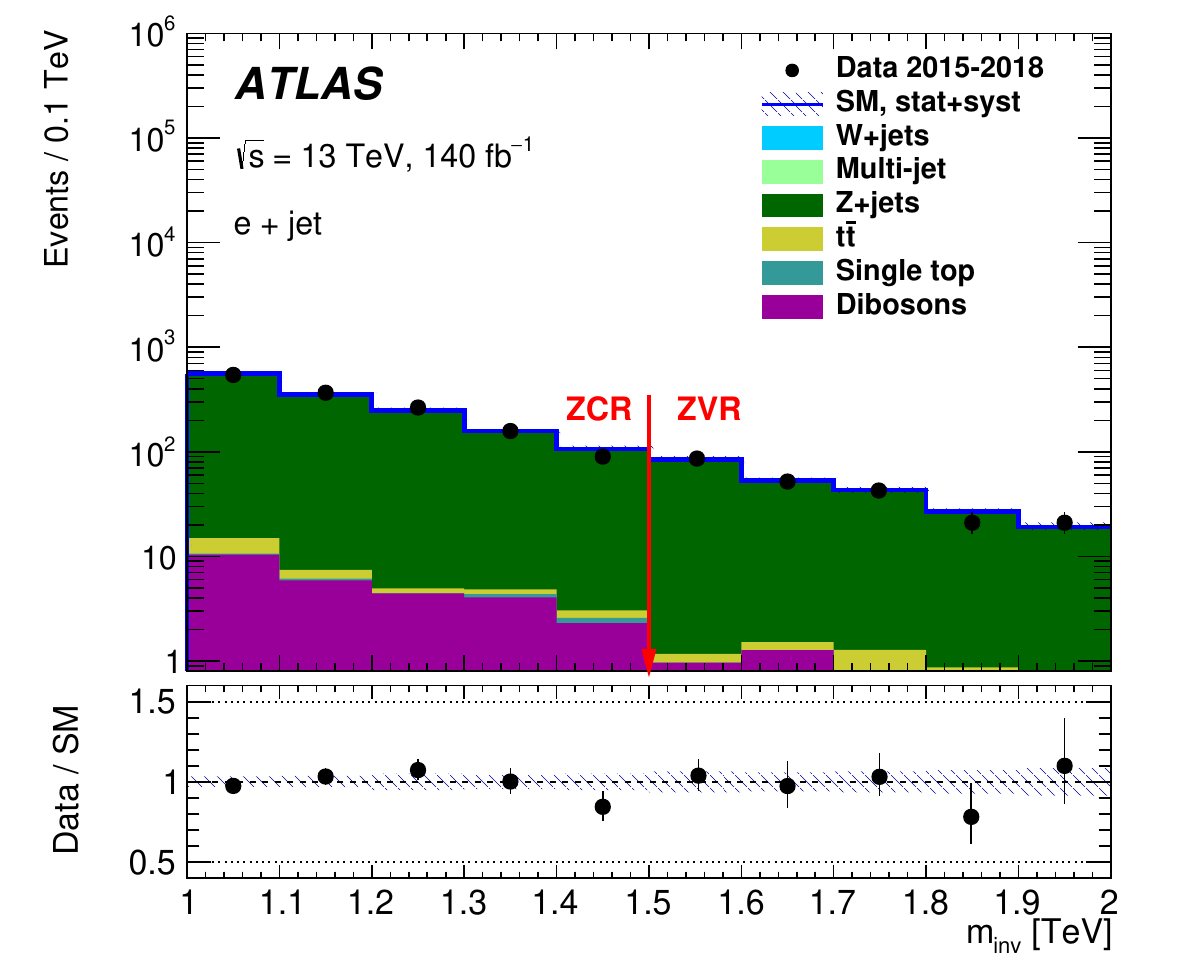}}
\subfloat[]{\includegraphics[width= 0.4\textwidth]{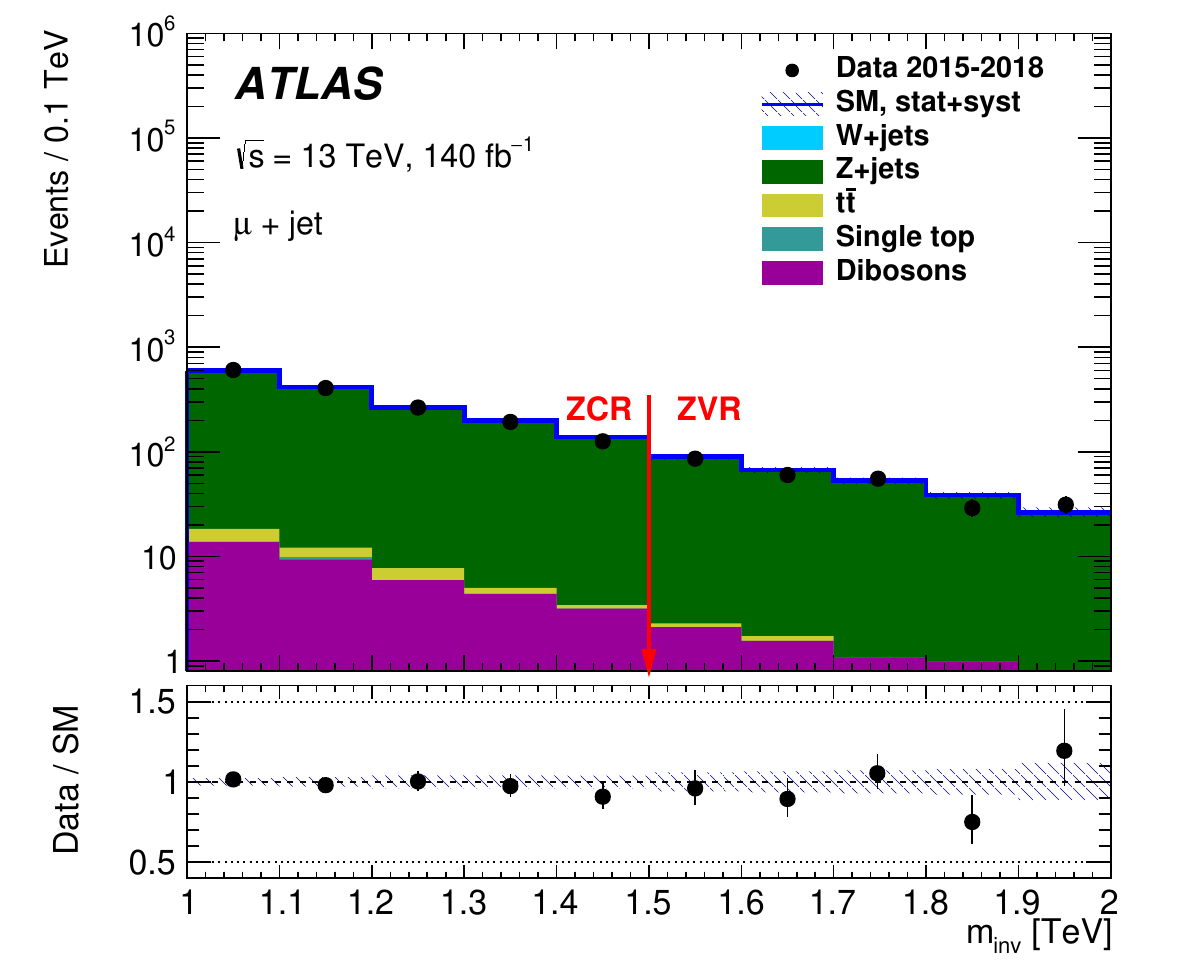}}\\
\subfloat[]{\includegraphics[width= 0.4\textwidth]{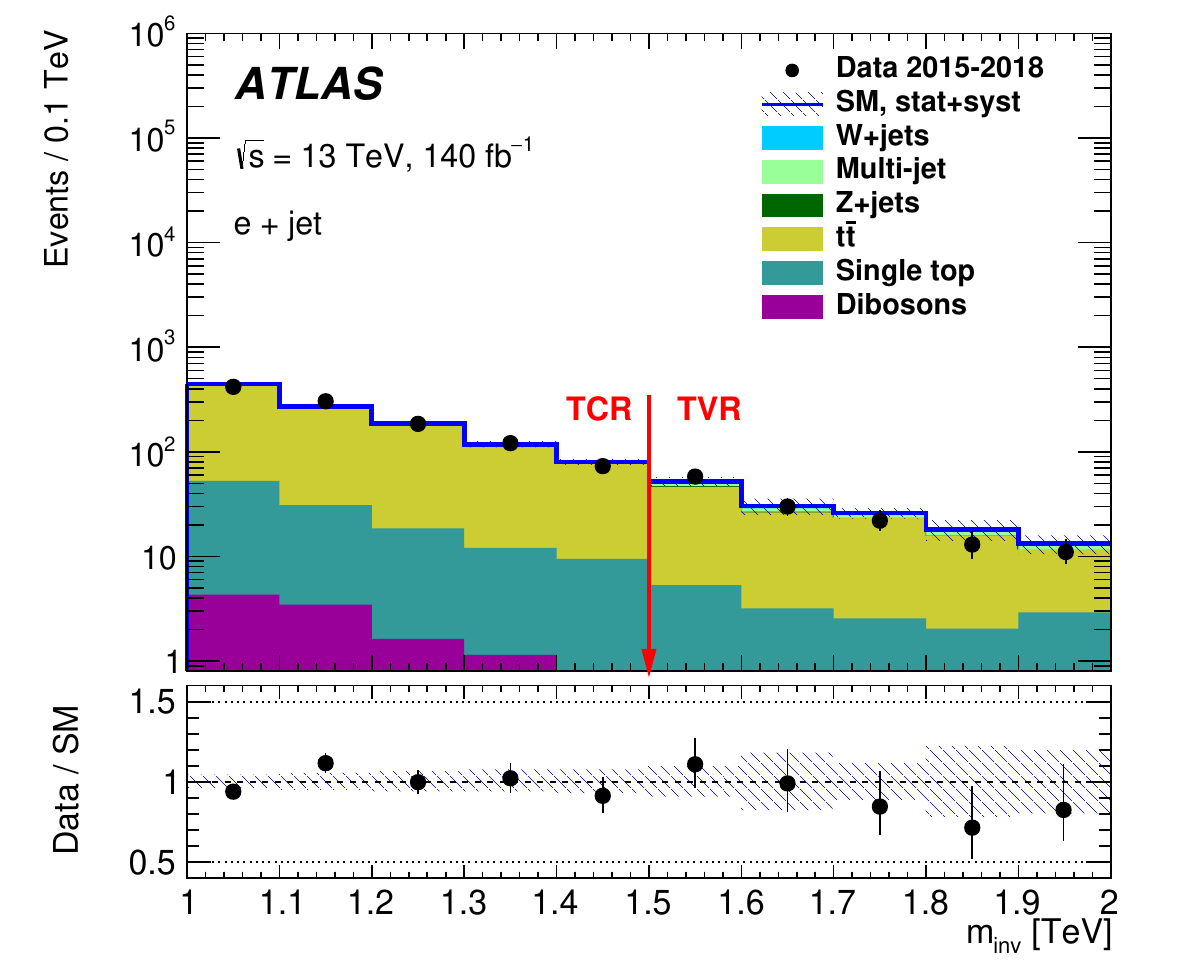}}
\subfloat[]{\includegraphics[width= 0.4\textwidth]{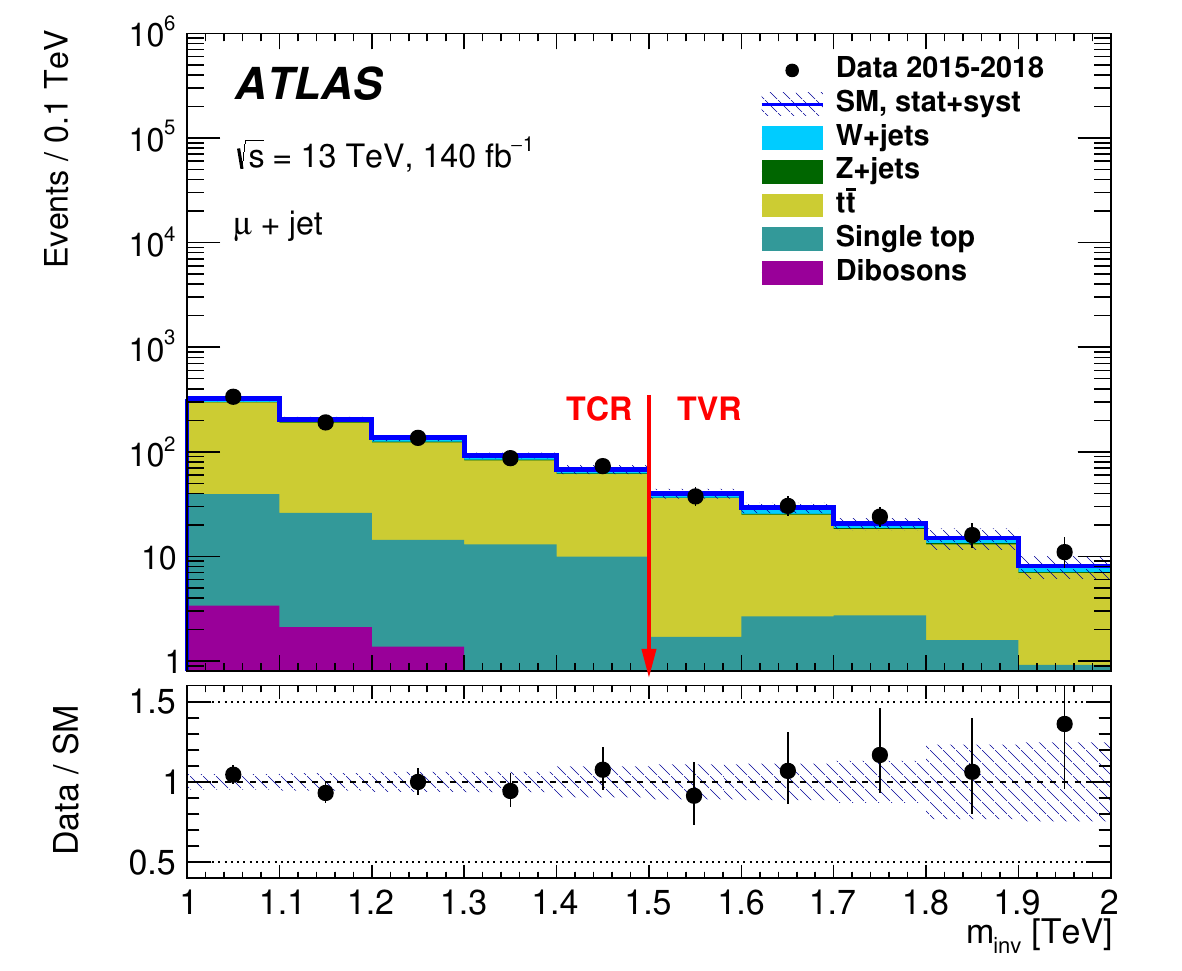}}\\
\caption{The distributions of events over the invariant mass of the leading lepton
and the leading jet are shown after the \textit{background-only} fit. The data
(points with error bars) and SM backgrounds (solid histograms) are shown in
(a), (c), and (e) for the electron+jet channel and in (b), (d), and (f) for the muon+jet channel.
The normalizations extracted from the fit in the CRs are applied in the full \Minv range.
(a), (b) The WCR and WVR. (c), (d) The ZCR and ZVR. (e), (f) The TCR and TVR.
The lower panels show the ratio of the number of events observed in the data to
the fitted total background. The hatched bands represent the total relative
uncertainty in the background estimate.}
\label{fig1}
\end{figure}

The \Minv distributions after the \textit{background-only} fit shown in Fig.~\ref{fig2}
have good agreement between the data and the SM background in the SR in both the electron+jet
and the muon+jet channels. The differences between the data and background are within 1$\sigma$.
The highest invariant mass of a lepton+jet pair reconstructed in the electron
(muon) channel is 4.74~\tev\ (4.96~\tev).

\begin{figure}[!htbp]
\centering
\subfloat[]{\includegraphics[width= 0.48\textwidth]{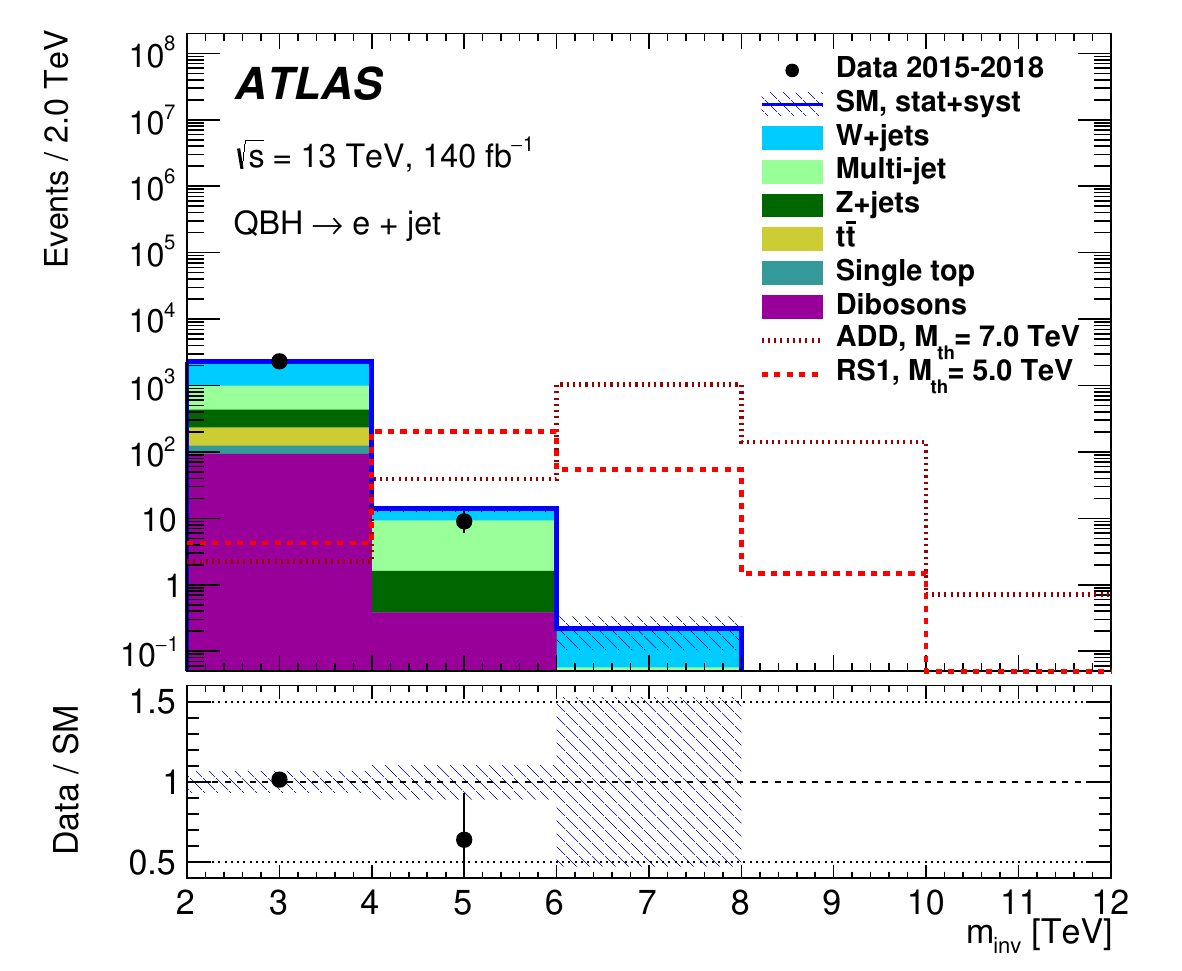}}
\subfloat[]{\includegraphics[width= 0.48\textwidth]{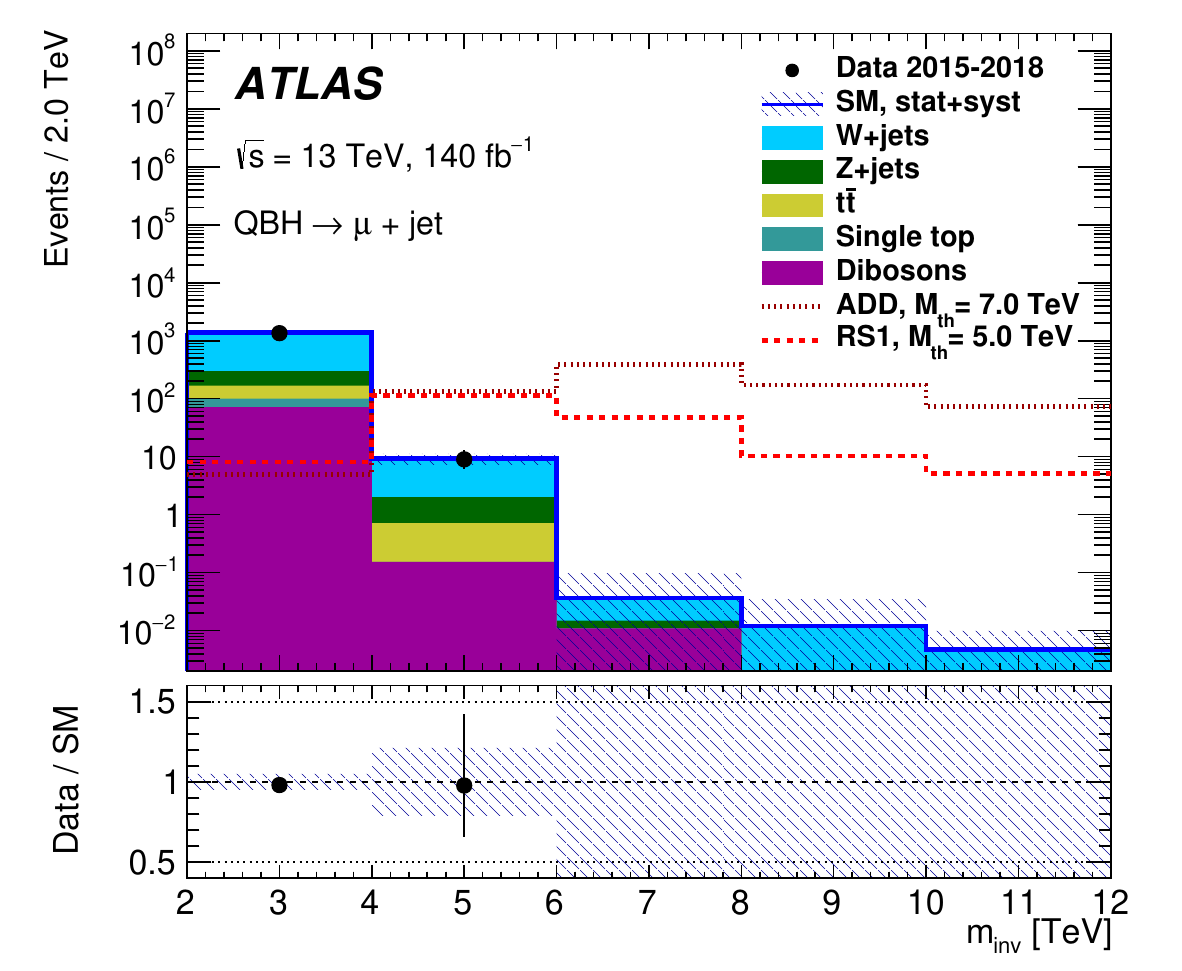}}\\
\caption{The distributions of events over the invariant mass of the leading lepton
and the leading jet in the SR for data (points with error bars) and for SM
backgrounds (solid histograms) after the \textit{background-only} fit are shown in
(a) the electron+jet and (b) the muon+jet channels. The normalizations extracted
from the fit in the CRs are applied in the full \Minv range including the SR.
The sum of the systematic uncertainties and the statistical errors due to
the limited size of the fitted MC samples is shown by the hatched area.
The lower panels show the ratios of the number of events
observed in the data to the fitted total background.
The hatched band represents the total relative uncertainty in the background estimate.
Two examples of QBH signals normalized to the predicted cross section are overlaid.}
\label{fig2}
\end{figure}

The \textit{model-independent} fit is performed simultaneously in the WCR, ZCR, TCR
and a single-bin SR to test for a non-SM signal contribution.
The possible contribution of signal events is scaled by a freely floating signal
normalization factor.
No significant excess above the SM background prediction is observed in either of the channels.
The \textit{model-independent} upper limit on the cross section times branching fraction
($\sigma\times Br$) is estimated at 95\%
confidence level (C.L.) for the production of a non-SM signal.

Figure~\ref{fig3} shows the upper limits on the $\sigma\times Br$ (circles along the
solid red line) integrated above the lower threshold of the SR (events with $m_{inv}>$ Th$_\mathrm{SR}$)
for the lepton+jet channel (combined channel of electron+jet and muon+jet).

\begin{figure}[!htbp]
\centering
\includegraphics[width=0.48\textwidth]{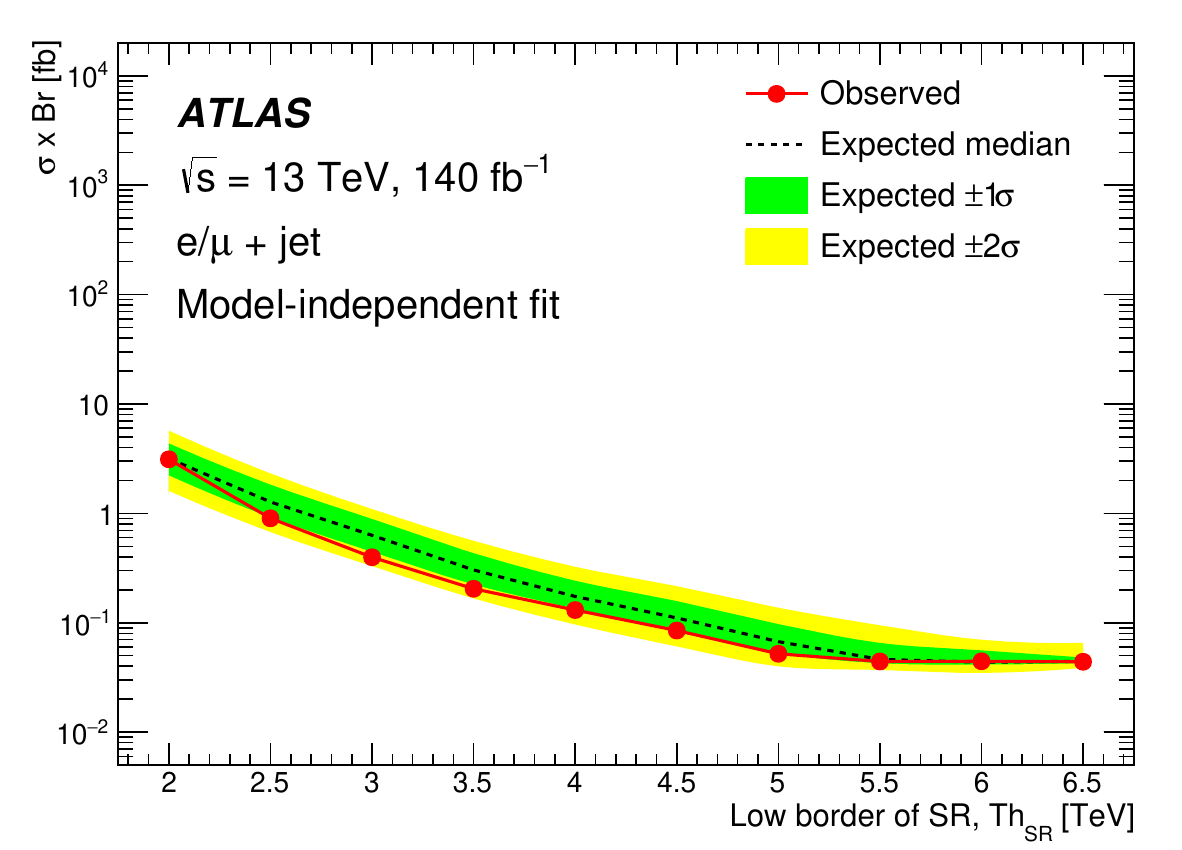}
\caption{The 95\% C.L. \textit{model-independent} upper limits on $\sigma \times Br$
for the non-SM signal production with decay into lepton+jet (combined channel).
The limits take into account statistical and systematic uncertainties.
Circles along the solid red line indicate the lower border of the SR (threshold of SR,
Th$_\mathrm{SR}$), above which the observed limit is computed.
The expected limits are shown by the dashed line. The $\pm 1\sigma$ and
$\pm 2\sigma$ bands of expected limits are shown in green and yellow, respectively.
The limits are obtained with pseudoexperiments.}
\label{fig3}
\end{figure}

In the \textit{model-dependent} fit, the 5-bin \Minv distributions of signal and backgrounds
in the SR are fitted simultaneously with background in three CRs.
The number of ADD (RS1) signal events is scaled by a freely floating signal normalization factor.
The background normalization factors are also determined simultaneously in the fit in the CRs,
and they are consistent with those of the \textit{background-only} fit.
There is no evidence of a QBH signal at any $M_\mathrm{th}$ in both models.
Figure~\ref{fig4} shows the 95\% C.L. upper limit on the cross section times branching
fraction\footnote{There are six QBH states that can decay to lepton+jet.
As each state has a different production cross section and branching fraction,
the limits set an effective limit which is a sum over all possible QBH states.}
($\sigma\times Br$) as a function of $M_\mathrm{th}$ for the combined lepton+jet
channel for the production of a QBH in the ADD and RS1 models.
The limits are obtained using pseudoexperiments with a spacing of
0.5~\tev\ in $M_\mathrm{th}$ and they are linearly interpolated between the points.

\begin{figure}[!htbp]
\centering
\subfloat[]{\includegraphics[width=0.48\textwidth]{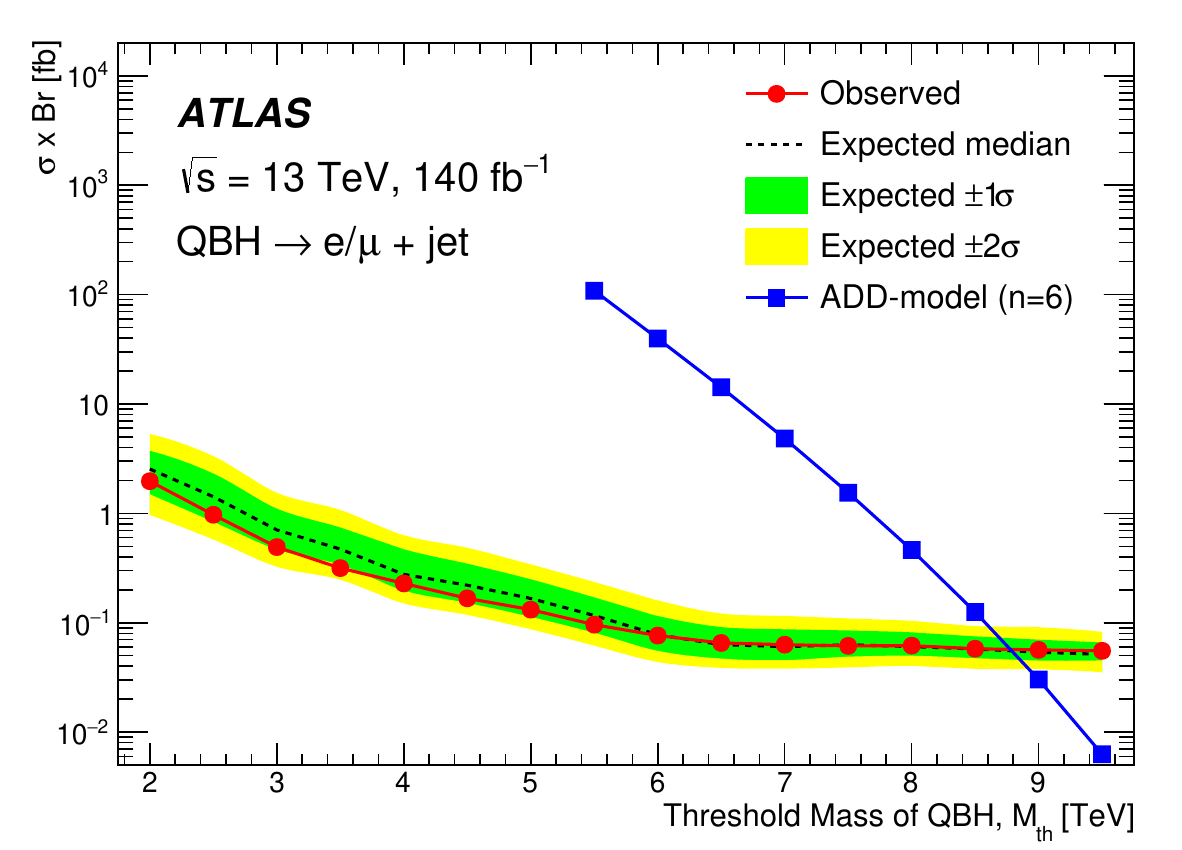}}
\subfloat[]{\includegraphics[width=0.48\textwidth]{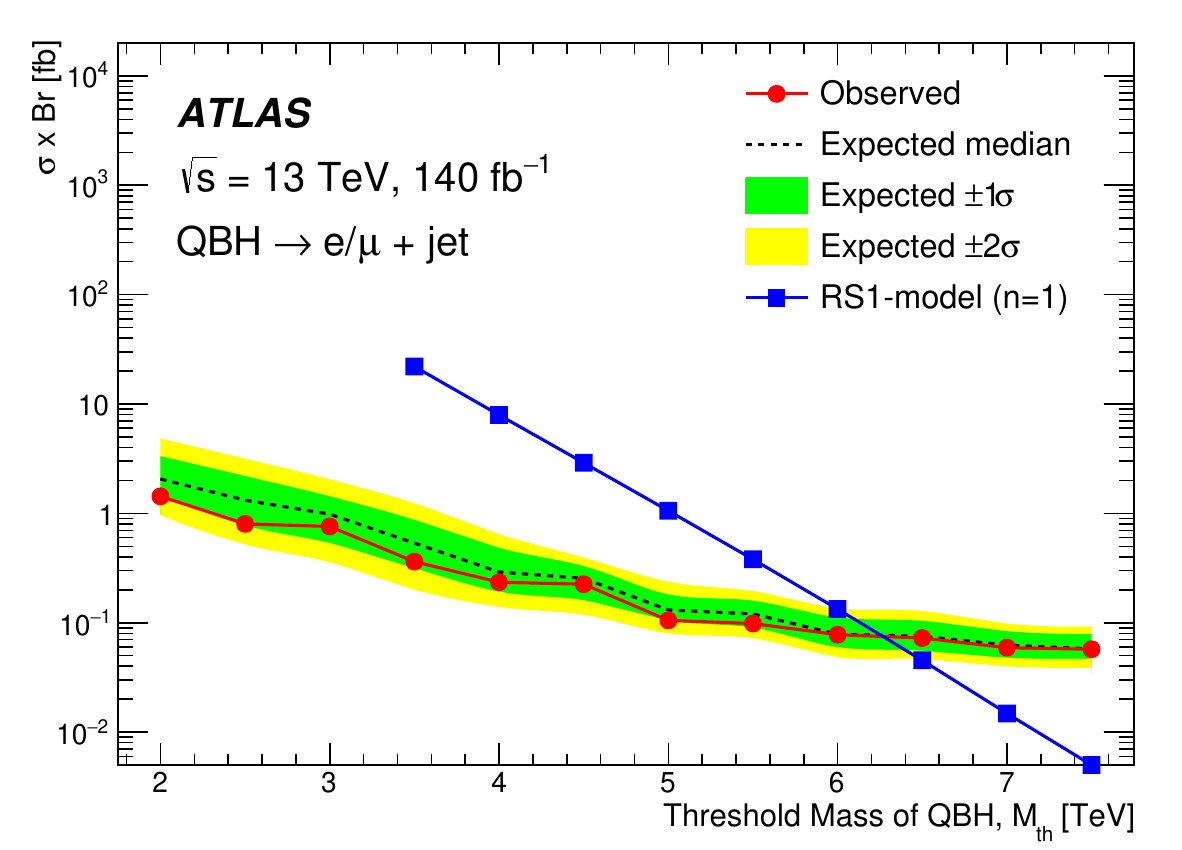}}\\
\caption{The combined 95\% C.L. upper limits on $\sigma \times Br$ as a function of $M_\mathrm{th}$
for QBH production at $M_\mathrm{th} = M_\mathrm{D}$ with decay into lepton+jet for
(a) ADD (extra dimensions $n = 6$) and (b) RS1 (extra dimensions $n = 1$).
The limits take into account statistical and systematic uncertainties.
Circles along the solid red line indicate the mass $M_\mathrm{th}$ of the signal
where the observed limit is computed. The expected limits
are shown by the dashed line. The $\pm 1\sigma$ and
$\pm 2\sigma$ bands are shown in green and yellow, respectively.
The theoretically predicted $\sigma \times Br$ for the QBH production and decay
is shown as the solid blue curve with squares.
The limits are obtained with pseudoexperiments.}
\label{fig4}
\end{figure}

The lower limits on $M_\mathrm{th}$ for ADD and RS1, upper limits on $\sigma\times Br$
at the $M_\mathrm{th}$ mass point limits and model-independent upper limits on
$\sigma(\Minv>5~\tev)\times Br$ are shown in Table~\ref{tab:UpLim}.
Accounting for QCD radiation effects in the QBH production using the $R_c$ correction factor
leads to more stringent limit estimates than without it. Future QBH lepton+jet analyses
have the potential to explore higher QBH mass ranges and lower QBH production cross section
values once hard QCD radiation effects are included in the QBH event generation model.

\begin{table}[htb]
\centering
\caption{The lower limits on $M_\mathrm{th}$ and the upper limits on $\sigma \times Br$
at these mass points for QBHs decaying to a lepton and jet in the ADD and RS1 models.
The model-independent upper limits on $\sigma \times Br$ are shown at \Minv > 5 TeV.}
\label{tab:UpLim}
\small
\begin{tabular}{lccccc}\hline\hline
Channel       &  ADD    &  ADD  &  RS1    &  RS1  &  Model-independent  \\
&  $\sigma\times Br$ [fb]   &  $M_\mathrm{th}$ [\tev]  &  $\sigma\times Br$ [fb]
&  $M_\mathrm{th}$ [\tev]   &  $\sigma(\Minv>5~\tev)\times Br$ [fb]  \\\hline
Electron+jet  &  0.091  &  8.6  &  0.100  &  6.1  &  0.095  \\
Muon+jet      &  0.083  &  8.6  &  0.090  &  6.2  &  0.084  \\
Combined      &  0.057  &  8.8  &  0.075  &  6.3  &  0.052  \\\hline\hline
\end{tabular}
\normalsize
\end{table}

\FloatBarrier


%
\section{CONCLUSION}
\label{sec:conclusion}

The ATLAS detector at the LHC has been used to search for new
phenomena in the lepton+jet invariant mass spectrum.
The search is performed with $140~\ifb$ of proton-proton collision
data at $\sqrt{s} = 13$~\TeV, recorded during 2015--2018.
The observed invariant mass spectrum of lepton+jet pairs is consistent
with SM expectations.
Upper exclusion limits are set on the cross section times branching
fraction for quantum black holes decaying to a lepton and a quark
in a search region with invariant  mass above 2.0~\TeV.
The resulting lower mass threshold limits
in the ADD (RS1) models with six (one) extra dimensions
at the 95\% C.L. are 8.8 (6.3)~\tev.
The obtained limits show a factor of 3.5 improvement with respect to the previous
model-independent upper limit on $\sigma\times Br$~\cite{EXOT-2013-03}.
The obtained limit on the QBH threshold mass for the ADD model is 3.5~\tev\ higher
compared to the previous ATLAS result in this channel at 8 TeV~\cite{EXOT-2013-03}.
The obtained limit on the QBH $M_\mathrm{th}$ for the RS1 model is determined
for the first time in the lepton+jet decay mode.


%

%
\section*{Acknowledgements}

%
%

%
%

We thank CERN for the very successful operation of the LHC, as well as the
support staff from our institutions without whom ATLAS could not be
operated efficiently.

We acknowledge the support of
ANPCyT, Argentina;
YerPhI, Armenia;
ARC, Australia;
BMWFW and FWF, Austria;
ANAS, Azerbaijan;
CNPq and FAPESP, Brazil;
NSERC, NRC and CFI, Canada;
CERN;
ANID, Chile;
CAS, MOST and NSFC, China;
Minciencias, Colombia;
MEYS CR, Czech Republic;
DNRF and DNSRC, Denmark;
IN2P3-CNRS and CEA-DRF/IRFU, France;
SRNSFG, Georgia;
BMBF, HGF and MPG, Germany;
GSRI, Greece;
RGC and Hong Kong SAR, China;
ISF and Benoziyo Center, Israel;
INFN, Italy;
MEXT and JSPS, Japan;
CNRST, Morocco;
NWO, Netherlands;
RCN, Norway;
MEiN, Poland;
FCT, Portugal;
MNE/IFA, Romania;
MESTD, Serbia;
MSSR, Slovakia;
ARRS and MIZ\v{S}, Slovenia;
DSI/NRF, South Africa;
MICINN, Spain;
SRC and Wallenberg Foundation, Sweden;
SERI, SNSF and Cantons of Bern and Geneva, Switzerland;
MOST, Taiwan;
TENMAK, T\"urkiye;
STFC, United Kingdom;
DOE and NSF, United States of America.
In addition, individual groups and members have received support from
BCKDF, CANARIE, Compute Canada and CRC, Canada;
PRIMUS 21/SCI/017 and UNCE SCI/013, Czech Republic;
COST, ERC, ERDF, Horizon 2020, ICSC-NextGenerationEU and Marie Sk{\l}odowska-Curie Actions, European Union;
Investissements d'Avenir Labex, Investissements d'Avenir Idex and ANR, France;
DFG and AvH Foundation, Germany;
Herakleitos, Thales and Aristeia programmes co-financed by EU-ESF and the Greek NSRF, Greece;
BSF-NSF and MINERVA, Israel;
Norwegian Financial Mechanism 2014-2021, Norway;
NCN and NAWA, Poland;
La Caixa Banking Foundation, CERCA Programme Generalitat de Catalunya and PROMETEO and GenT Programmes Generalitat Valenciana, Spain;
G\"{o}ran Gustafssons Stiftelse, Sweden;
The Royal Society and Leverhulme Trust, United Kingdom.

The crucial computing support from all WLCG partners is acknowledged gratefully, in particular from CERN, the ATLAS Tier-1 facilities at TRIUMF (Canada), NDGF (Denmark, Norway, Sweden), CC-IN2P3 (France), KIT/GridKA (Germany), INFN-CNAF (Italy), NL-T1 (Netherlands), PIC (Spain), ASGC (Taiwan), RAL (UK) and BNL (USA), the Tier-2 facilities worldwide and large non-WLCG resource providers. Major contributors of computing resources are listed in Ref.~\cite{ATL-SOFT-PUB-2023-001}.

%
%


\clearpage
\printbibliography
\clearpage
 
\begin{flushleft}
\hypersetup{urlcolor=black}
{\Large ATLAS Collaboration}

\bigskip

\AtlasOrcid[0000-0002-6665-4934]{G.~Aad}$^\textrm{\scriptsize 103}$,
\AtlasOrcid[0000-0002-5888-2734]{B.~Abbott}$^\textrm{\scriptsize 121}$,
\AtlasOrcid[0000-0002-7248-3203]{D.C.~Abbott}$^\textrm{\scriptsize 104}$,
\AtlasOrcid[0000-0002-2788-3822]{A.~Abed~Abud}$^\textrm{\scriptsize 36}$,
\AtlasOrcid[0000-0002-1002-1652]{K.~Abeling}$^\textrm{\scriptsize 55}$,
\AtlasOrcid[0000-0002-2987-4006]{D.K.~Abhayasinghe}$^\textrm{\scriptsize 95}$,
\AtlasOrcid[0000-0002-8496-9294]{S.H.~Abidi}$^\textrm{\scriptsize 158}$,
\AtlasOrcid[0000-0002-8279-9324]{O.S.~AbouZeid}$^\textrm{\scriptsize 42}$,
\AtlasOrcid{N.L.~Abraham}$^\textrm{\scriptsize 148}$,
\AtlasOrcid[0000-0001-5329-6640]{H.~Abramowicz}$^\textrm{\scriptsize 153}$
% ...
\bigskip
\\

% ...
$^{36}$CERN, Geneva, Switzerland\\
% ...
$^{42}$Niels Bohr Institute, University of Copenhagen, Copenhagen, Denmark\\
% ...
$^{55}$II. Physikalisches Institut, Georg-August-Universit\"{a}t G\"ottingen, G\"ottingen, Germany\\
% ...
$^{95}$Department of Physics, Royal Holloway University of London, Egham, United Kingdom\\
% ...
$^{103}$CPPM, Aix-Marseille Universit\'e, CNRS/IN2P3, Marseille, France\\
$^{104}$Department of Physics, University of Massachusetts, Amherst, Massachusetts, USA\\
% ...
$^{121}$Homer L. Dodge Department of Physics and Astronomy, University of Oklahoma, Norman, Oklahoma, USA\\
% ...
$^{148}$Department of Physics and Astronomy, University of Sussex, Brighton, United Kingdom\\
% ...
$^{153}$Raymond and Beverly Sackler School of Physics and Astronomy, Tel Aviv University, Tel Aviv, Israel\\
% ...
$^{158}$Department of Physics, University of Toronto, Toronto, Ontario, Canada\\
% ...

\end{flushleft}


%
%
%
%
%

\end{document}